\documentclass{elsart}

\usepackage[square,comma]{natbib}
\usepackage{graphicx}
\usepackage{pxfonts}
\usepackage{lineno}

\usepackage{amssymb}

\journal{}

\begin{document}

\thispagestyle{empty}
\begin{Large}
\textbf{DEUTSCHES ELEKTRONEN-SYNCHROTRON}

\textbf{\large{Ein Forschungszentrum der
Helmholtz-Gemeinschaft}\\}
\end{Large}

DESY 10-199

November 2010

\begin{eqnarray}
\nonumber &&\cr \nonumber && \cr \nonumber &&\cr
\end{eqnarray}
\begin{eqnarray}
\nonumber
\end{eqnarray}
\begin{center}
\begin{Large}
\textbf{Generation of doublet spectral lines at self-seeded X-ray FELs}
\end{Large}
\begin{eqnarray}
\nonumber &&\cr \nonumber && \cr
\end{eqnarray}

\begin{large}
Gianluca Geloni,
\end{large}
\textsl{\\European XFEL GmbH, Hamburg}
\begin{large}

Vitali Kocharyan and Evgeni Saldin
\end{large}
\textsl{\\Deutsches Elektronen-Synchrotron DESY, Hamburg}
\begin{eqnarray}
\nonumber
\end{eqnarray}
\begin{eqnarray}
\nonumber
\end{eqnarray}
ISSN 0418-9833
\begin{eqnarray}
\nonumber
\end{eqnarray}
\begin{large}
\textbf{NOTKESTRASSE 85 - 22607 HAMBURG}
\end{large}
\end{center}
\clearpage
\newpage

\begin{frontmatter}



\title{Generation of doublet spectral lines at self-seeded X-ray FELs}


\author[XFEL]{Gianluca Geloni\thanksref{corr},}
\thanks[corr]{Corresponding Author. E-mail address: gianluca.geloni@xfel.eu}
\author[DESY]{Vitali Kocharyan}
\author[DESY]{and Evgeni Saldin}

\address[XFEL]{European XFEL GmbH, Hamburg, Germany}
\address[DESY]{Deutsches Elektronen-Synchrotron (DESY), Hamburg,
Germany}

\begin{abstract}
Self-seeding schemes, consisting of two undulators with a monochromator in between, aim to reduce the bandwidth of SASE X-ray FELs.  We recently proposed to use a new method of monochromatization exploiting a single crystal in Bragg-transmission geometry for self-seeding in the hard X-ray range. The obvious and technically possible extension is to use such kind of monochromator setup with two -or more- crystals arranged in a series to spectrally filter the SASE radiation at two -or more- closely-spaced wavelengths within the FEL gain band. This allows for the production of doublet -or multiplet- spectral lines. Applications exist over a broad range of hard X-ray wavelengths involving any process where there is a large change in cross section over a narrow wavelength range, as in multiple wavelength anomalous diffraction techniques (MAD). In this paper we consider the simultaneous operation of the LCLS hard X-ray FEL at two closely spaced wavelengths. We present simulation results for the LCLS baseline, and we show that this method can produce fully coherent radiation shared between two longitudinal modes. Mode spacing can be easily tuned within the FEL gain band, i.e. within $10$ eV. An interesting aspect of the proposed scheme is a way of modulating the electron bunch at optical frequencies without a seed quantum laser. In fact, the XFEL output intensity contains an oscillating "mode-beat" component whose frequency is related to the frequency difference between the pair of longitudinal modes considered.  Thus, at saturation one obtains FEL-induced modulations of energy loss and energy spread in the electron bunch at optical frequency. These modulations can be converted into density modulation at the same optical frequency with the help of a weak chicane installed behind the baseline undulator. Powerful coherent radiation can then be generated with the help of an optical transition radiation (OTR) station, which have important applications. In this paper we briefly consider how the doublet structure of the XFEL generation spectra can be monitored by an optical spectrometer. Furthermore, the OTR coherent radiation pulse is naturally synchronized with the X-ray pulses, and can be used for timing the XFEL to high power conventional lasers with femtosecond accuracy for pump-probe applications.
\end{abstract}

%
%
%
\end{frontmatter}



\section{\label{sec:intro} Introduction}

LCLS began routine user operation with photon energy up to $10$ keV \cite{LCLS2}. This success motivated the planning of a significant upgrade over the next several years \cite{FRIS}. Plans for the near-term upgrade include self-seeding for hard X-rays \cite{FRIS}.

In \cite{OURY4}-\cite{OURY5}, self-seeding options for the LCLS baseline were investigated using a scheme relying on a single-crystal monochromator in Bragg-transmission geometry. The Bragg crystal reflects a narrow band of X-rays, resulting in a ringing within the passband in the forward direction, which can be used to seed the second undulator. The chicane creates a transverse offset for the electrons, washes out previous electron-beam microbunching, and provides a tunable delay of the electron-bunch with respect to the radiation, so that the electron beam only interacts with the ringing tail of the X-ray pulse \cite{OURX}-\cite{OURY3}.

In this paper we extend our previous investigations to the case of a self-seeding scheme with a similar kind of monochromator, consisting of two or more crystals in the Bragg geometry arranged in a series. The impinging SASE radiation is therefore spectrally filtered at two or more closely spaced wavelengths within the FEL gain band. This allows for generation of doublet spectral lines.

Simultaneous operation of a high-gain FEL at two separate wavelengths using narrow band seed lasers was previously studied in  \cite{FREU}. Applications of such mode of operation involve any process where there is a large change in cross section over narrow wavelength range, as in multiple wavelength anomalous diffraction technique  \cite{MADH}. Here we present simulation results for the LCLS baseline and we show that this method can produce a doublet structure in the LCLS hard X-ray FEL spectrum.

The XFEL-output intensity contains an oscillating "mode-beat" component whose frequency is related to the frequency difference between the pair of longitudinal modes considered. With the help of a weak chicane installed behind the baseline undulator, see \cite{BROA}, a powerful, coherent pulse of radiation can then be generated with the help of an optical transition radiation (OTR) station, which has important applications. In this paper we briefly consider how the generation of doublet structures in the XFEL spectrum can be monitored by an optical spectrometer. Furthermore, the OTR coherent radiation pulse is naturally synchronized with the X-ray pulses, and can be used for timing the XFEL to high power conventional lasers with femtosecond accuracy for pump-probe applications.

\section{\label{sec:oper} Operation of the LCLS hard X-ray FEL at closely spaced wavelengths by use of a self-seeding scheme with a wake monochromator}

\begin{figure}[tb]
\includegraphics[width=1.0\textwidth]{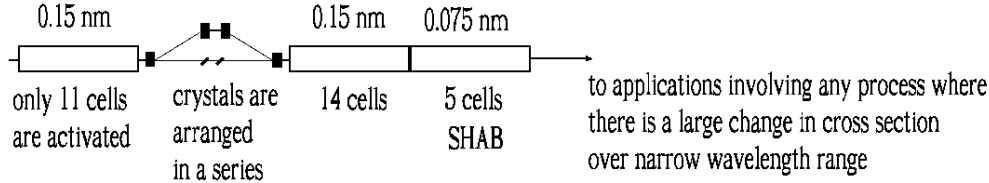}
\caption{Design of the LCLS baseline undulator system for generation of doublet spectral lines.} \label{lclsd3}
\end{figure}

In the present Section we introduce our technique. An overview is sketched in Fig. \ref{lclsd3}. The reader may recognize similarities with the previously proposed schemes \cite{OURY4}-\cite{OURY3}, except for the presence of two crystals, instead of one, located in the transverse offset provided by the magnetic chicane. As before, the first undulator in Fig. \ref{lclsd3} operates in the linear high-gain regime starting from the shot-noise in the electron beam.

\begin{figure}[tb]
\includegraphics[width=1.0\textwidth]{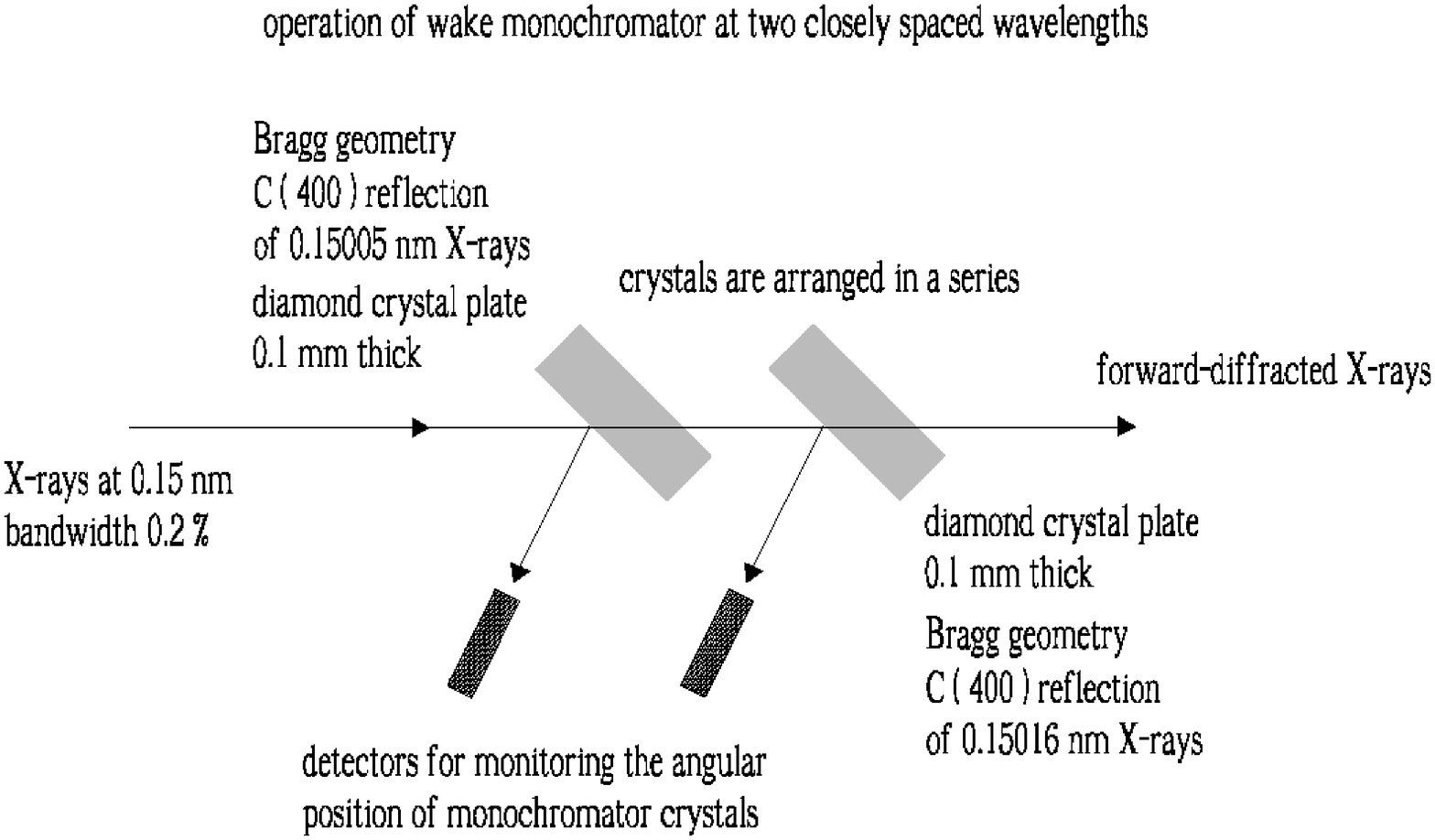}
\caption{Forward diffraction in a series of crystals in Bragg geometry. The incident angles on the first and on the second crystal are different. Due to multiple scattering, the transmittance spectrum of the crystals shows doublet absorption lines.  The temporal waveform
of the transmitted radiation pulse is characterized by a long tail. The radiation power within such tail is shared between
two harmonic waves with slightly different wavenumbers.} \label{lclsd1}
\end{figure}

\begin{figure}[tb]
\includegraphics[width=1.0\textwidth]{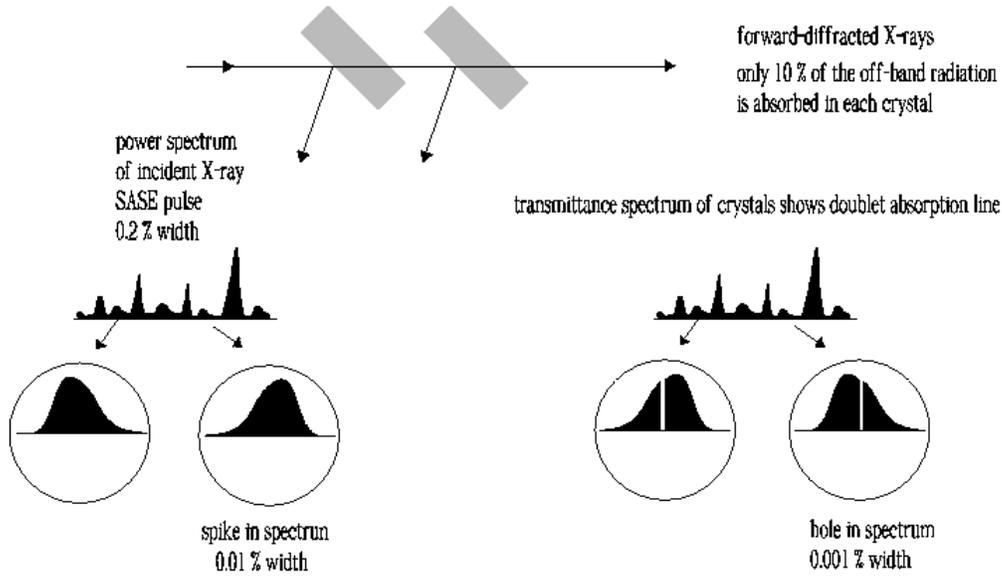}
\caption{Two crystals bandstop filter as monochromator for the self-seeding scheme. The two crystals in Bragg geometry are arranged in a series, and act as bandstop filters for the transmitted X-ray SASE radiation pulse. Each crystal reflects a narrow band of X-rays, resulting in a ringing in the pass band that can be used to seed the second undulator at two closely spaced wavelengths.} \label{lclsd2}
\end{figure}

\begin{figure}[tb]
\includegraphics[width=1.0\textwidth]{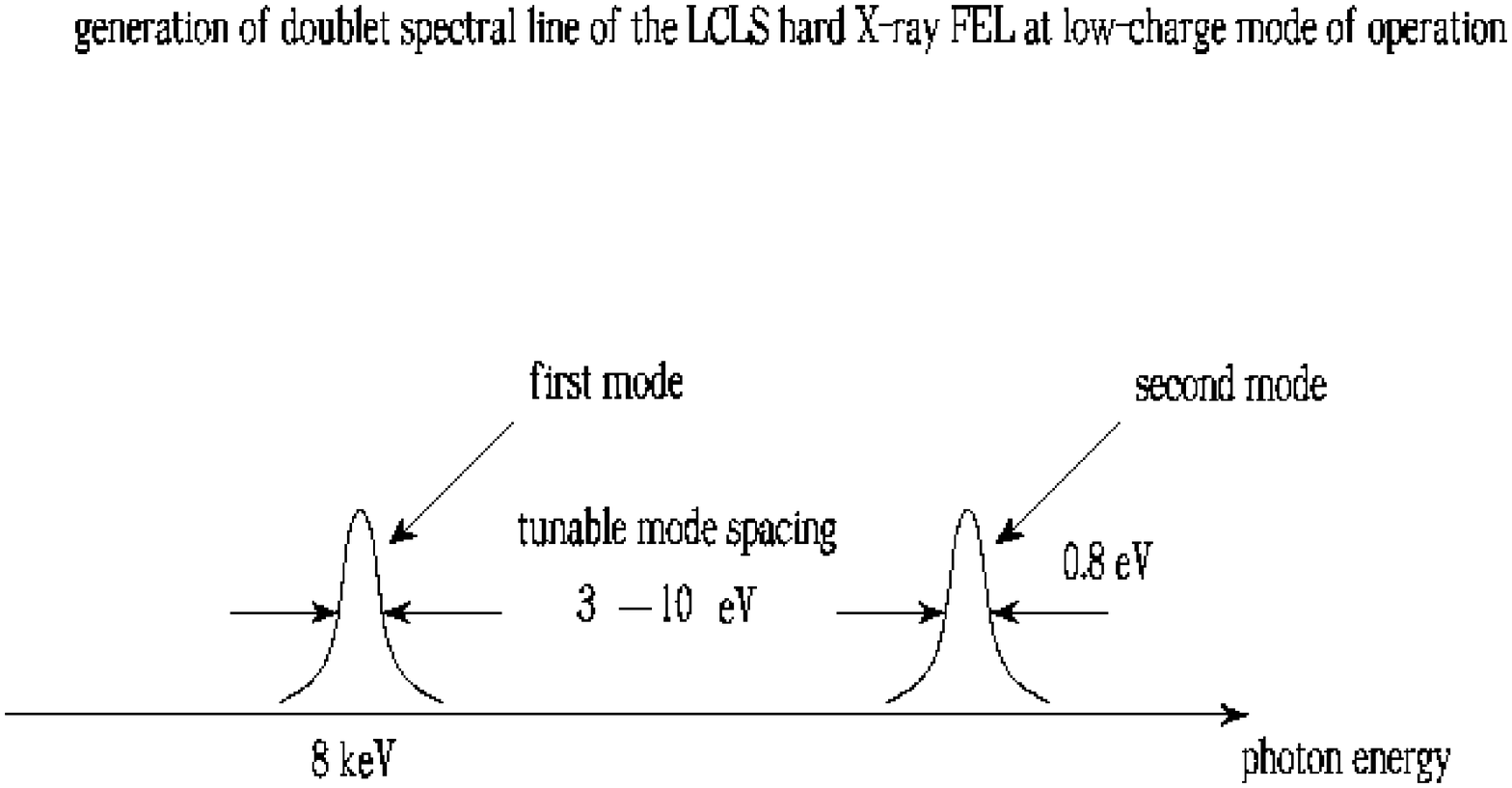}
\caption{The doublet structure of the X-ray FEL generation spectra.} \label{lclsd4}
\end{figure}

After the first undulator, the output SASE radiation passes through the monochromator, consisting of a series of two (or more) crystals in the Bragg transmission geometry. The monochromator setup is sketched in Fig. \ref{lclsd1}. According to our monochromator principle, the  SASE pulse coming from the first undulator impinges on a set of two or more crystals in Bragg diffraction geometry. Then, the crystals  operate as bandstop filters for the transmitted X-ray SASE radiation pulse, as shown in Fig. \ref{lclsd2}.  The incident angles of the first and of the second crystal are different, so that the Bragg condition is met for different frequencies within the SASE spectrum. The temporal waveform of the transmitted radiation pulse shows a long tail, whose duration is inversely proportional to the bandwidth of the absorption line in the transmittance spectrum. This tail can be used for seeding the electron bunch after the chicane. It should be noted here that the ringing tail does not depend on the distance between the crystals, and that such distance has no influence on the output characteristics of the radiation. At variance with previously considered schemes relying on a single crystal, here two bandstop filters are present at different frequencies. As a result, both frequency components are present in the seeding signal, and will be amplified in the output undulator. This allows for the production of doublet spectral lines, Fig. \ref{lclsd4}. The output signal from our setup is thus fully coherent, but shared between two separate longitudinal modes. The relative alignment tolerance of the crystal tilting angles should now be in the order of ten microrad, in order to allow for a stable frequency difference within the Darwin width.

While the radiation is sent through the crystals, the electron beam passes through a magnetic chicane, which accomplishes three tasks: it creates an offset for the crystals installation, it removes the electron microbunching produced in the first undulator, and it acts as a delay line for the implementation of a temporal windowing process. In this process, the magnetic chicane shifts the electron bunch on top of the monochromatic tail created by the bandstop filter thus temporally selecting a part of it. By this, the electron bunch is seeded with a radiation pulse characterized by a bandwidth much narrower than the natural FEL bandwidth. For the hard X-ray wavelength range, a small dispersive strength $R_{56}$ in the order of a few microns is sufficient to remove the microbunching generated in the first undulator part.  As a result, the choice of the strength of the magnetic chicane only depends on the delay that one wants to introduce between electron bunch  and radiation. The optimal value amounts to $R_{56} \simeq 12 ~\mu$m for the low-charge mode of operation. Such dispersion strength is small enough to be generated by a short ($4$ m-long) magnetic chicane to be installed in place of a single undulator module. Such chicane is, at the same time, strong enough to create a sufficiently large transverse offset for installing the crystals.

\begin{figure}[tb]
\includegraphics[width=0.5\textwidth]{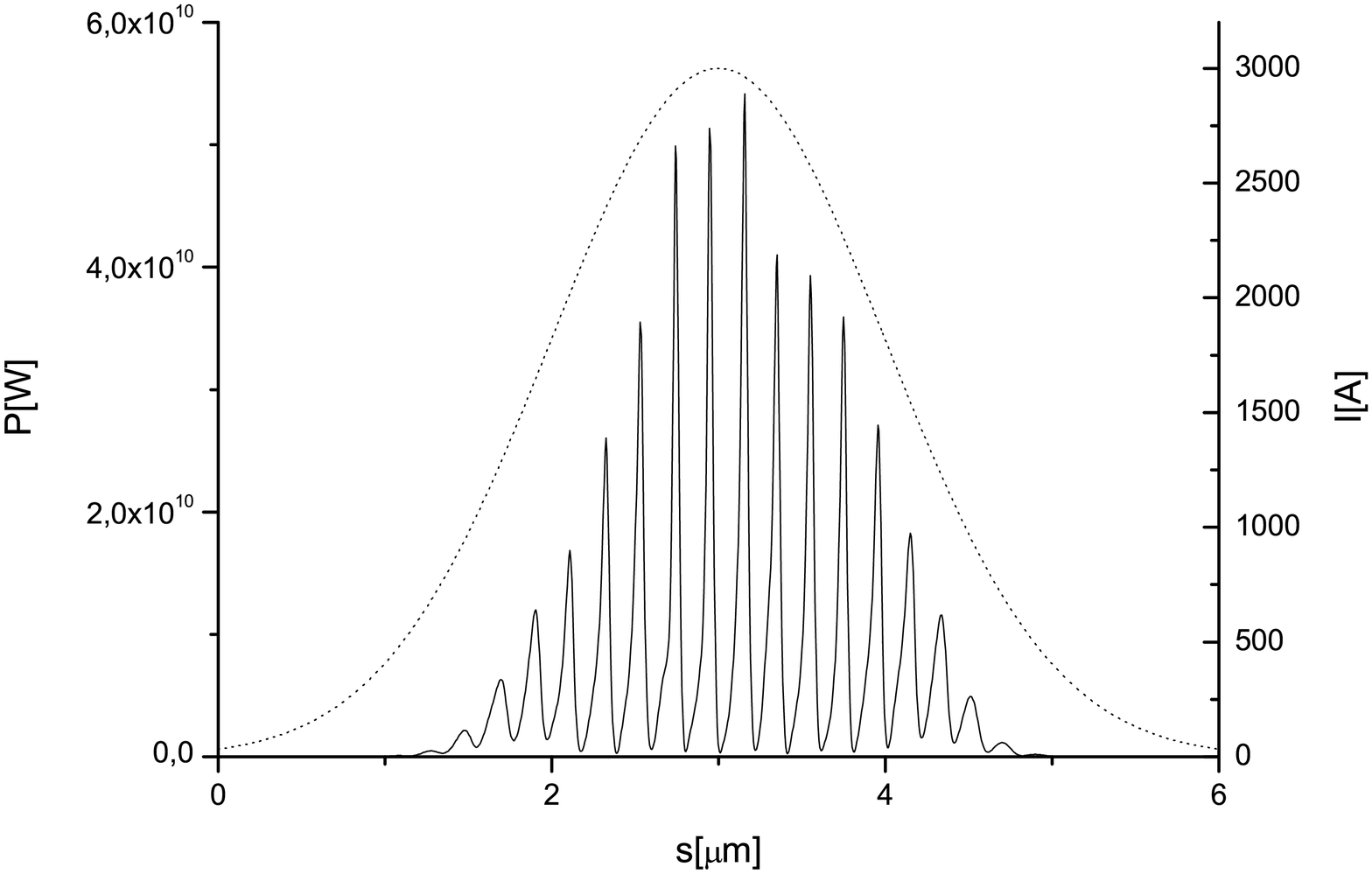}
\includegraphics[width=0.5\textwidth]{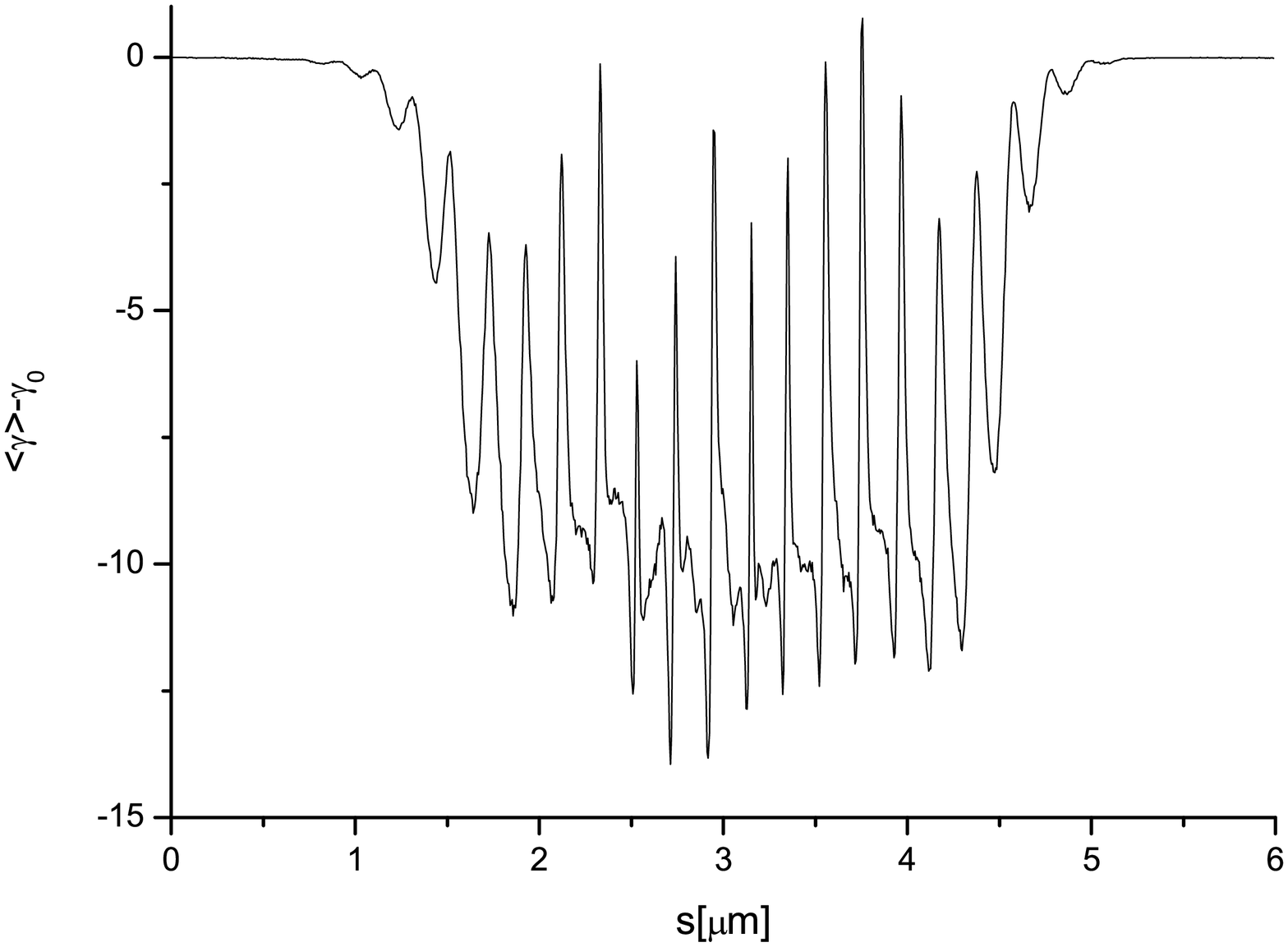}
\includegraphics[width=0.5\textwidth]{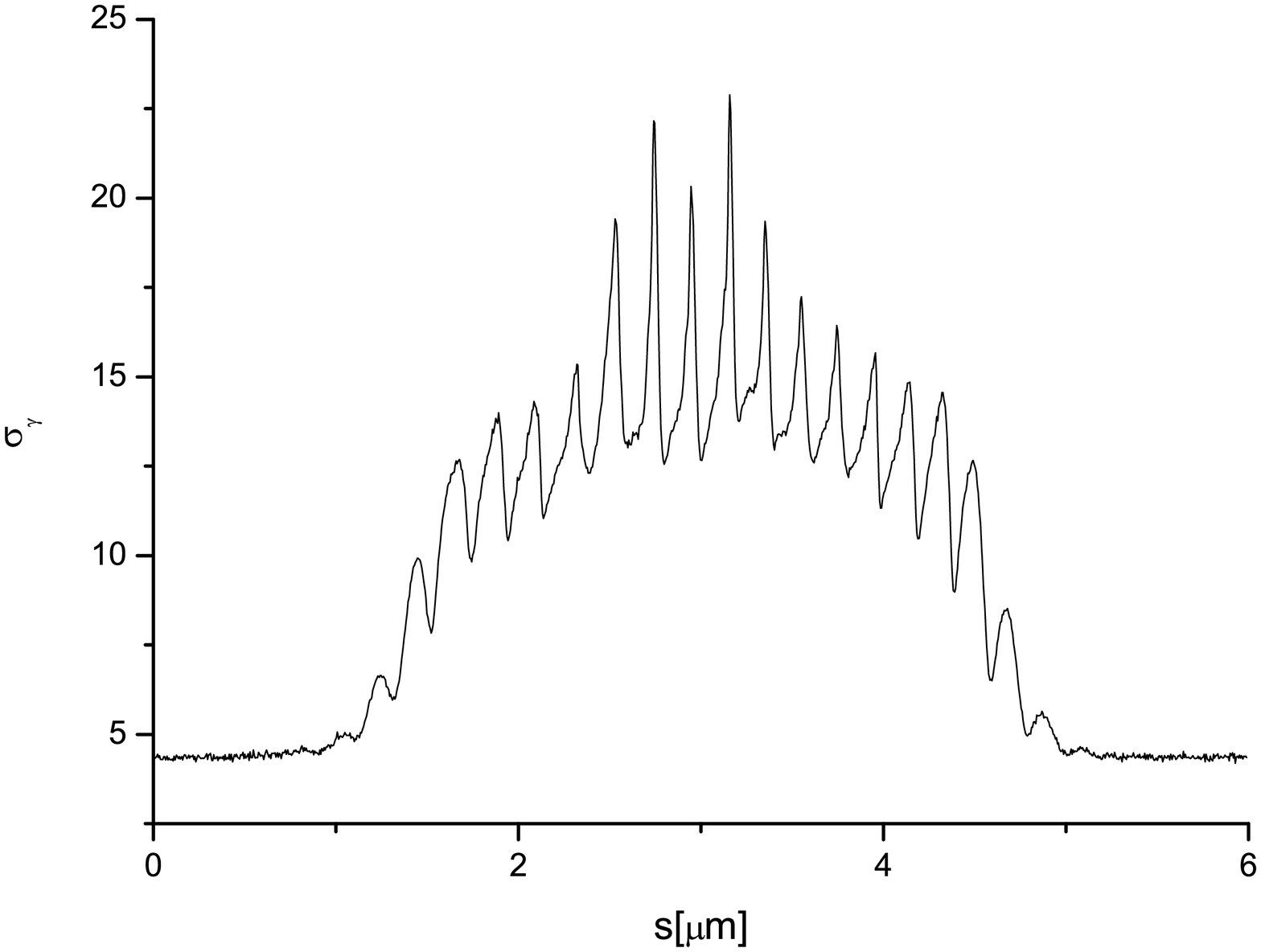}
\includegraphics[width=0.5\textwidth]{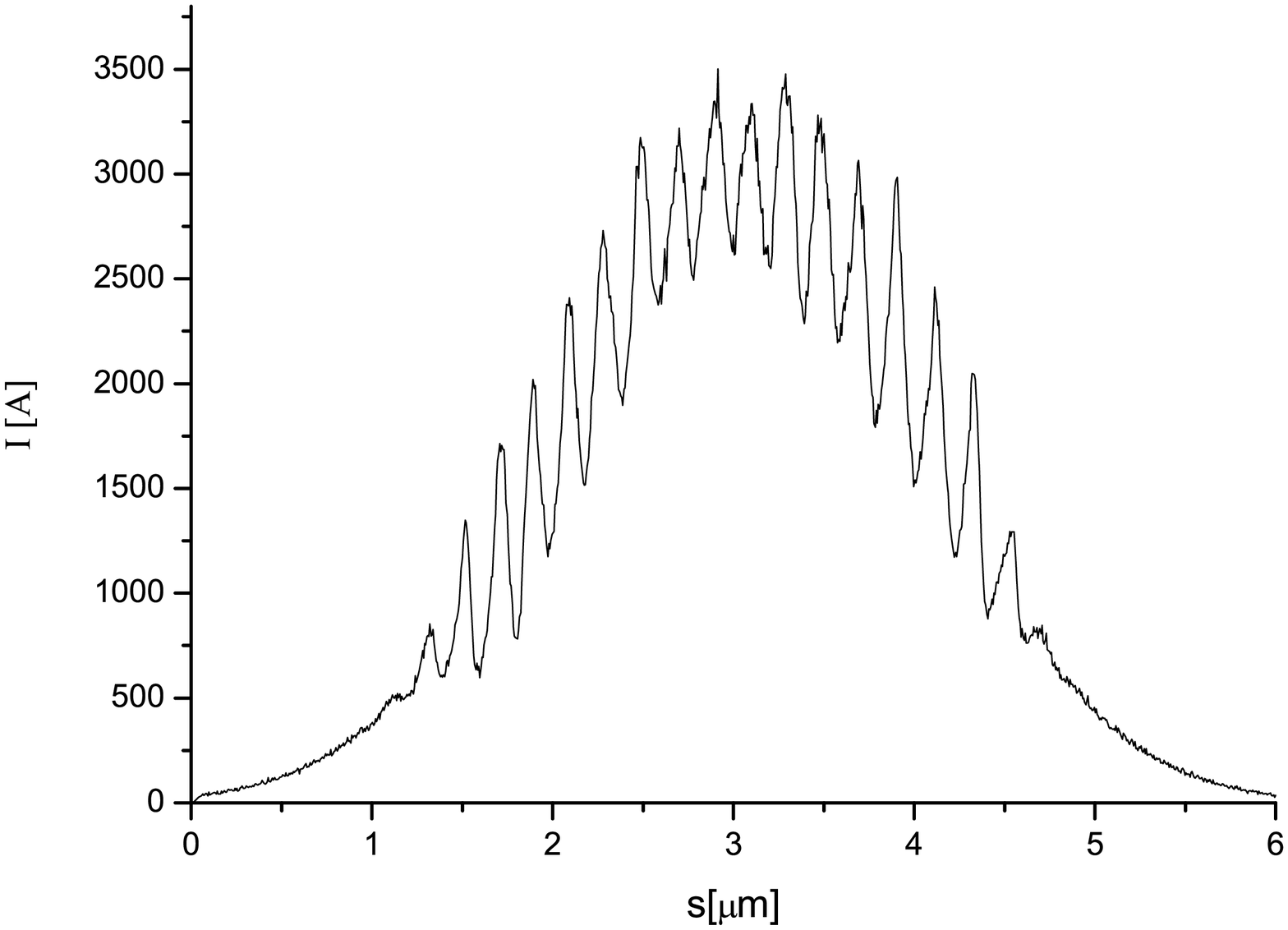}
\caption{Sample of out put data as a function of the position inside the bunch. (Upper left) Output power after $14$ undulator cells. The dotted line represents the electron bunch profile. (Upper right) Average energy loss. (Bottom left) rms energy spread. (Bottom right) Current profile after a weak chicane installed after the LCLS baseline undulator, $R_{56}=0.08$ mm.} \label{sampledat}
\end{figure}

Since two bandstop filters are present at slightly different frequencies, the seed signal contains a "mode-beat" component whose frequency is related to the difference between central frequencies of the bandstop filters. As a result, the output from our setup exhibits a remarkable feature: namely we have modulation of the output radiation power and of the energy loss and energy spread of the electron beam at a frequency related with the difference between the central frequencies of the bandstop filters. This is exemplified in Fig. \ref{sampledat}, upper right, upper left and bottom left plots, obtained with the code Genesis \cite{GENE}. In the example case shown in Fig. \ref{sampledat}, this modulation is around $200$ nm.

\begin{figure}[tb]
\includegraphics[width=1.0\textwidth]{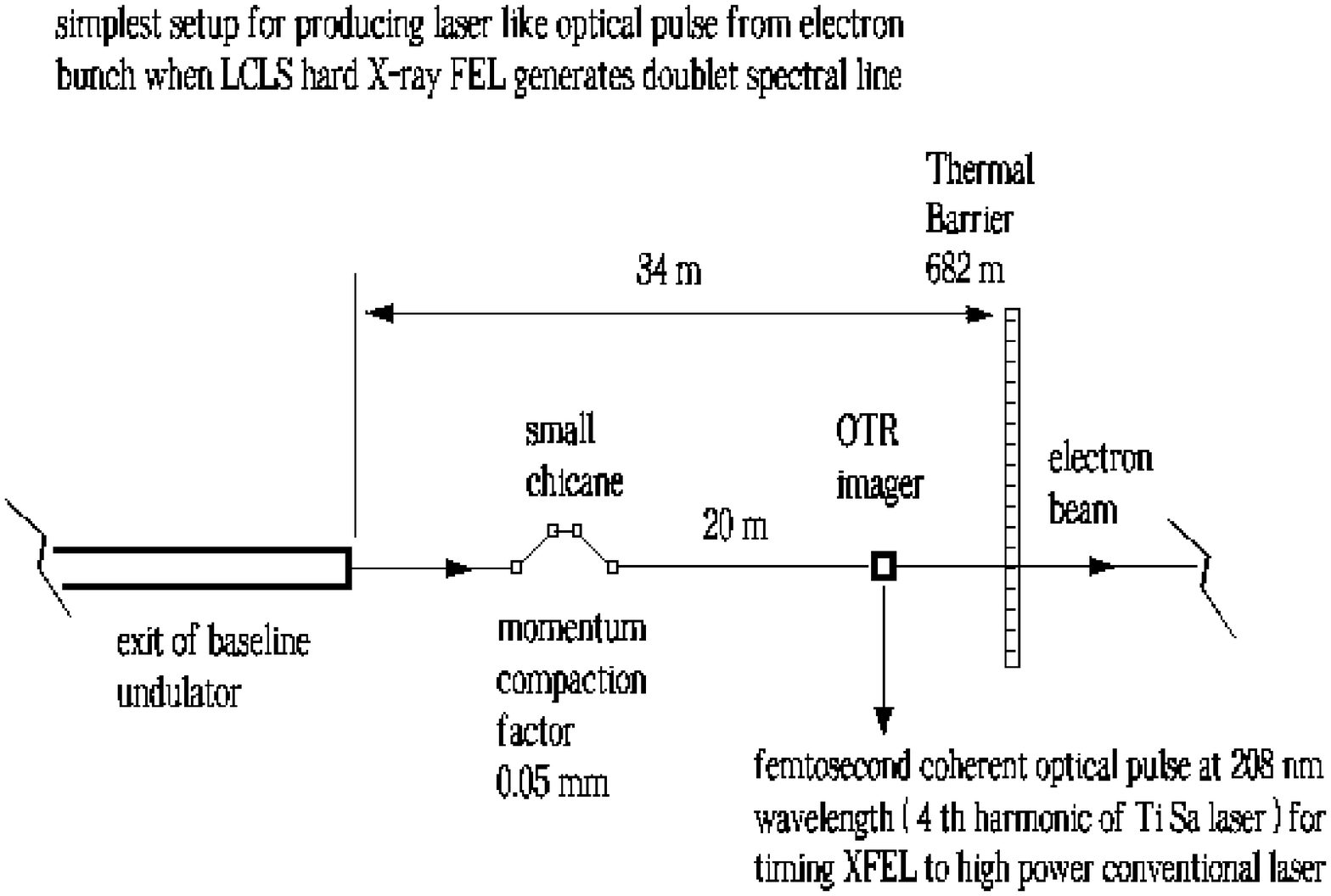}
\caption{Installation of a small dispersive element (magnetic chicane) and of an OTR station after the LCLS baseline undulator will allow to produce laser-like optical pulses from electron bunches when the LCLS hard X-ray FEL generates doublet spectral lines.} \label{lclsd5}
\end{figure}

One may take advantage of such energy modulation by transforming it into a density modulation, which can be used to produce powerful pulses of coherent radiation in the visible range. This can be done with the help of an extra magnetic chicane installed after the XFEL undulator at the LCLS baseline, as depicted in Fig. \ref{lclsd5}. A dispersive strength of $50-100$ microns is enough to this end. The output particle file from Genesis \cite{GENE} is used as input to the code Elegant \cite{ELEG} to study the electron beam dynamics through the magnetic chicane. Tracking results are shown in Fig. \ref{sampledat} bottom right.

\begin{figure}[tb]
\includegraphics[width=1.0\textwidth]{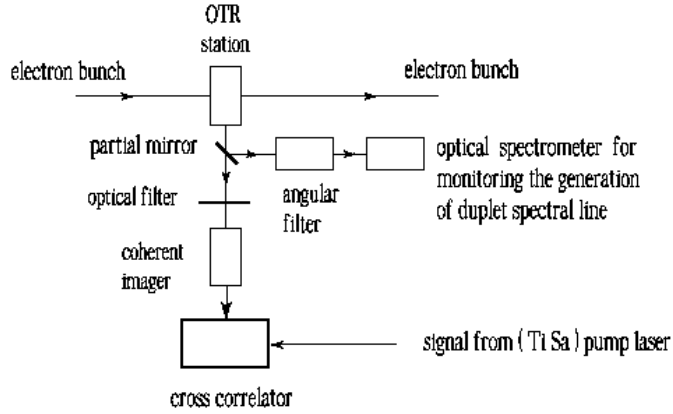}
\caption{Block diagram view of possible coherent OTR pulse applications.} \label{lclsd6}
\end{figure}

As a simple example of radiator one may consider an OTR station, as sketched in Fig. \ref{lclsd5} and Fig. \ref{lclsd6}. It is technically possible to avoid interference between the OTR pulse production and the X-ray beam delivery. In fact, the vacuum chamber downstream of the main undulator has an effective aperture of $25$ mm, which is the diameter of the BPM bore. Such aperture is sufficiently large to alter the electron beam trajectory (with the help of correctors) and introduce an offset in the order of a centimeter, thus separating the electron beam and the X-ray. Since we deal with optical wavelengths, such electron orbit perturbation  will not perturb the density modulation in electron bunch.

Several applications can be considered. With reference to Fig. \ref{lclsd6} the presence of coherent pulses of OTR can be used, with the help of an optical spectrometer, to monitor the generation of the doublet spectral lines. Moreover, the OTR pulses are naturally synchronized with the electron bunches, and therefore with the main X-ray FEL pulses. After filtering, these pulses can be cross-correlated with a harmonic signal from a TiSa pump laser in order to provide a timing of pump and probe pulses with an accuracy within a few femtoseconds.


\newpage

\section{\label{sec:FEL} FEL simulations}

Following the previous introduction to our technique, in this Section we report on a feasibility study performed with the help of the FEL code GENESIS 1.3 \cite{GENE} running on a parallel machine. In this Section we will present the feasibility study for the short-pulse mode of operation of the LCLS. Parameters used in the simulations for the short pulse mode of operation are presented in Table \ref{tt1}. We present a statistical analysis consisting of $100$ runs.

\begin{table}
\caption{Parameters for the low-charge mode of operation at LCLS used in
this paper.}

\begin{small}\begin{tabular}{ l c c}
\hline & ~ Units &  ~ \\ \hline
Undulator period      & mm                  & 30     \\
K parameter (rms)     & -                   & 2.466  \\
Wavelength            & nm                  & 0.15   \\
Energy                & GeV                 & 13.6   \\
Charge                & nC                  & 0.02 \\
Bunch length (rms)    & $\mu$m              & 1    \\
Normalized emittance  & mm~mrad             & 0.4    \\
Energy spread         & MeV                 & 1.5   \\
\hline
\end{tabular}\end{small}
\label{tt1}
\end{table}

\begin{figure}[tb]
\includegraphics[width=1.0\textwidth]{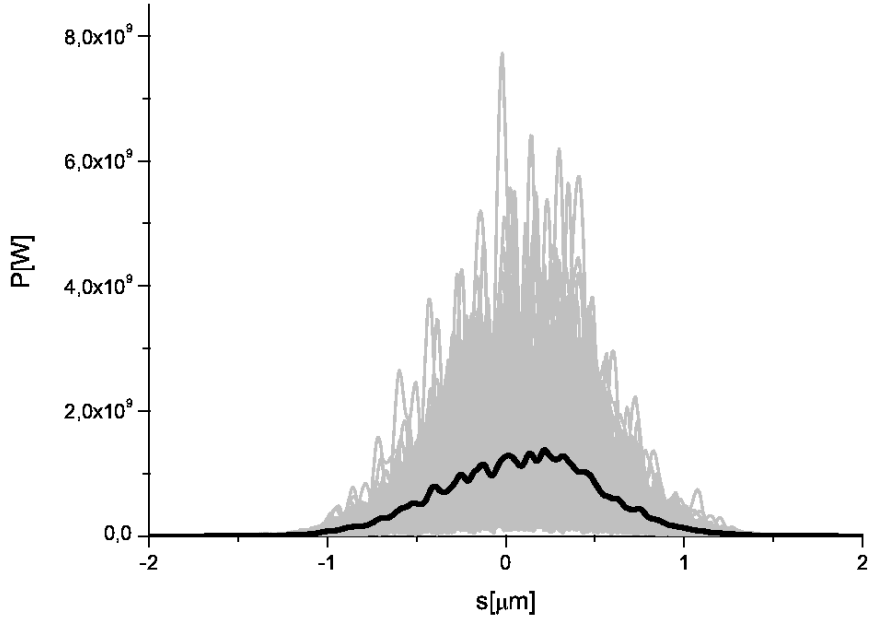}
\caption{Power distribution after the SASE undulator (11 cells). Grey lines refer to single shot realizations, the black line refers to the average over a hundred realizations.} \label{pin}
\end{figure}
\begin{figure}[tb]
\includegraphics[width=1.0\textwidth]{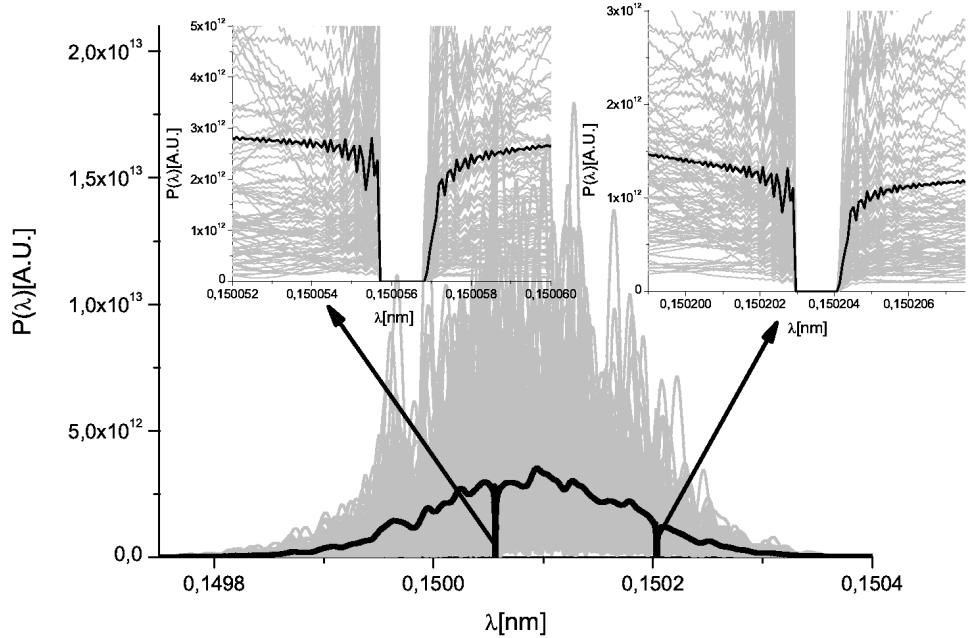}
\caption{Spectrum after the diamond crystals, $100~\mu$m-thick, C(400) reflection. The bandstop effect is clearly visible. Grey lines refer to single shot realizations, the black line refers to the average over a hundred realizations.} \label{spholes}
\end{figure}
\begin{figure}[tb]
\includegraphics[width=0.5\textwidth]{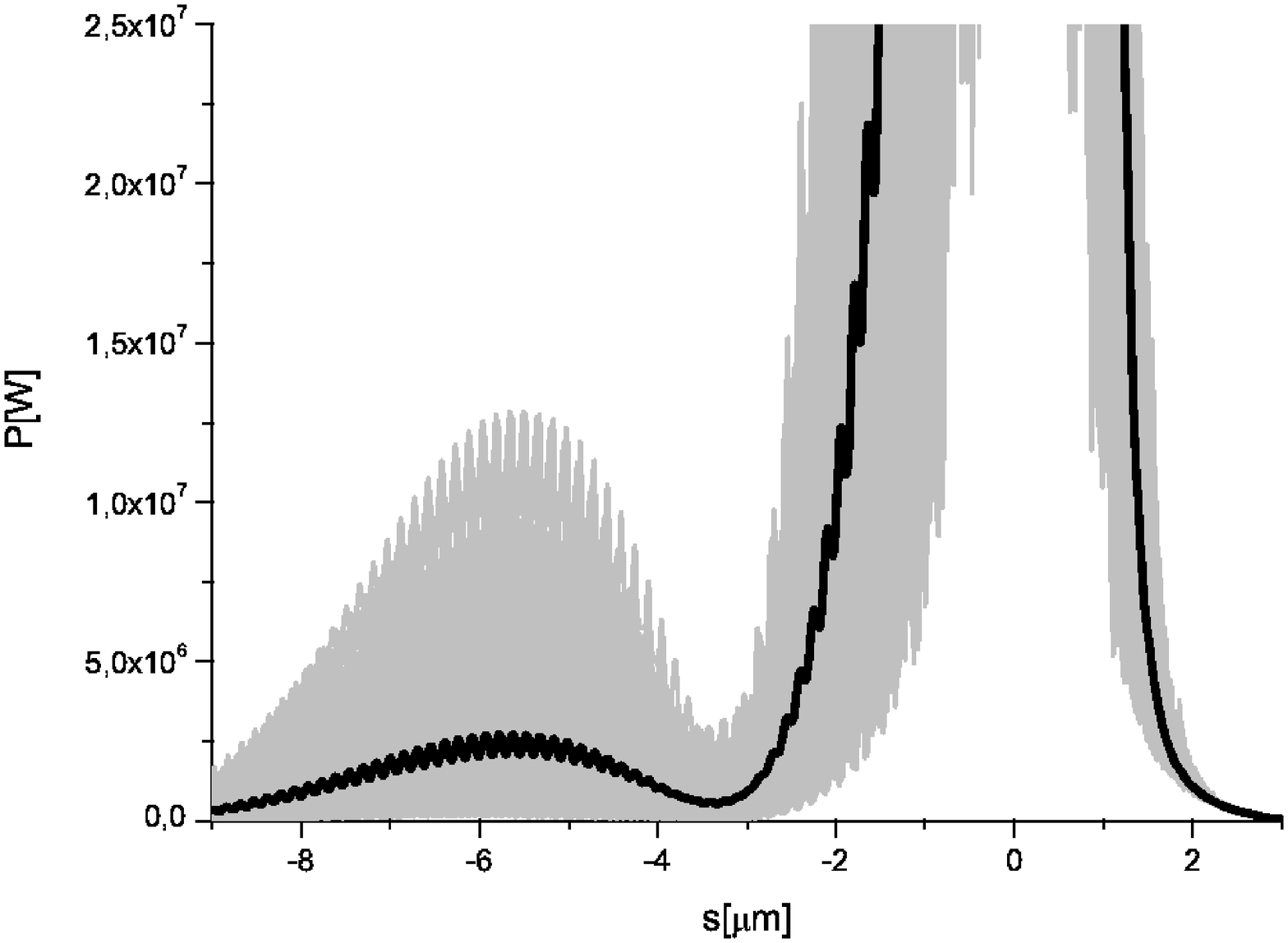}
\includegraphics[width=0.5\textwidth]{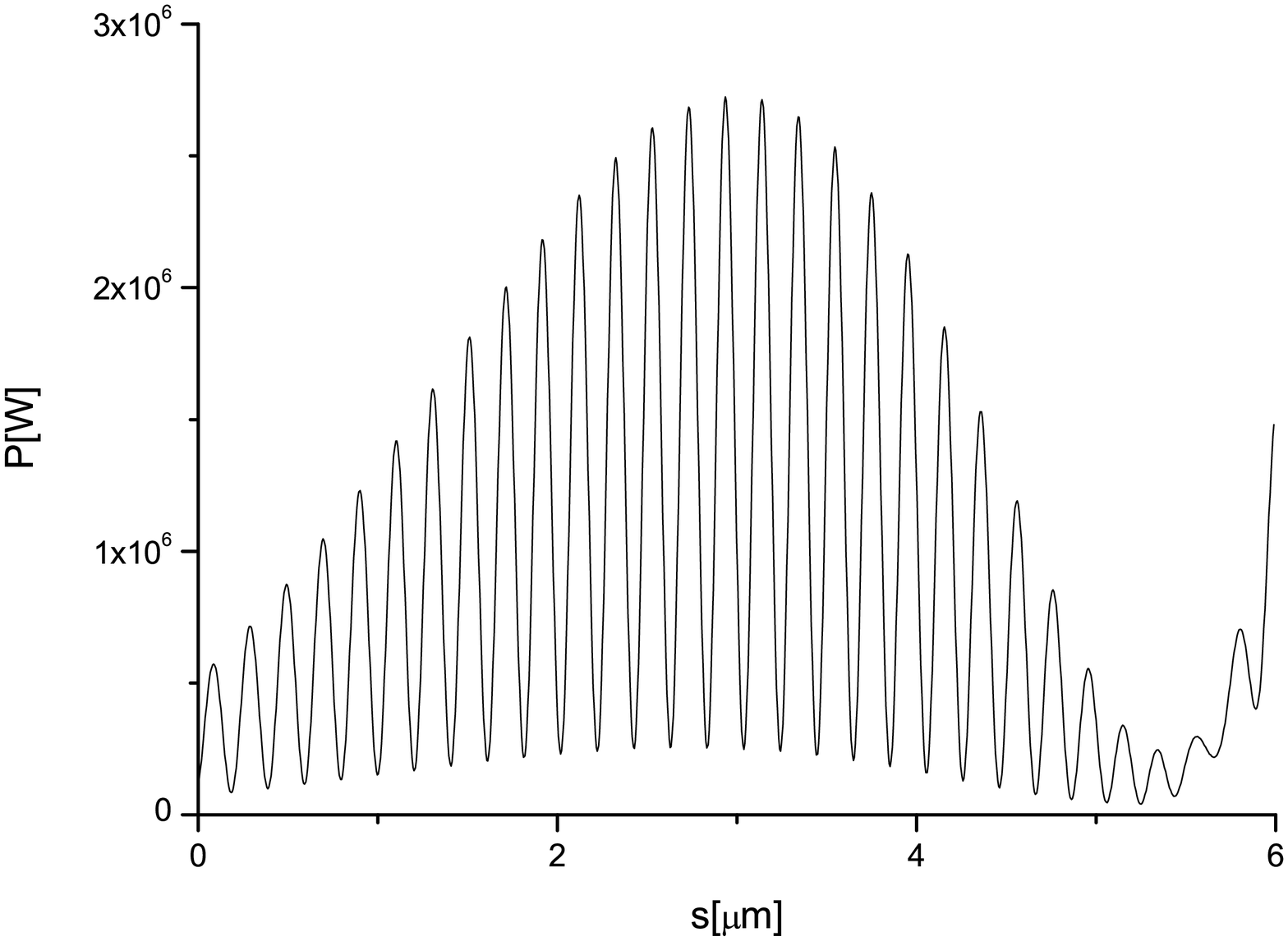}
\caption{(left) Power distribution after the diamond crystals. The monochromatic tail due to the transmission through the bandstop filters is now evident on the left of the figure. Grey lines refer to single shot realizations, the black line refers to the average over a hundred realizations. (right) An enlargement of a single shot seed power distribution presented in the left  part of the plot. The seed signal consists of  two harmonic waves with slightly different wavenumbers. The form of this type of signal is interesting, and consists of a rapidly oscillating (at X-ray angular frequency) carrier wave multiplied by a simple sinusoidal envelope function  at optical angular frequency. The beat of the sinusoidal envelope (see the maximum at 6 $\mu$m) is related to specific features of the transmittance profile of the band-stop filter.} \label{pseed}
\end{figure}

Our starting point is the simulation of the SASE radiation characteristics at the double crystal. The incident power distribution after $11$ section is shown in Fig. \ref{pin}. The bandstop-effect of the two diamond crystals (C(400) reflection, $100~\mu$m-thick crystal) is evident by inspection of Fig. \ref{spholes}. The effect in the time domain is shown in Fig. \ref{pseed}. As shown before in \cite{OURY4}-\cite{OURY5}, the presence of a single bandstop-filter is related with a long, monochromatic tail following the main SASE pulse. Such structure has a duration comparable with the inverse of the filter bandwidth. Moreover, since there are now two frequency filters centered at different central frequencies, a beating between the two waves at slightly different frequencies takes place, yielding a seeding power distribution modulated at optical frequencies, as is evident from the right part of Fig. \ref{pseed}, which is an enlargement of the average (black) line in the left part of the same figure.

\begin{figure}[tb]
\includegraphics[width=0.5\textwidth]{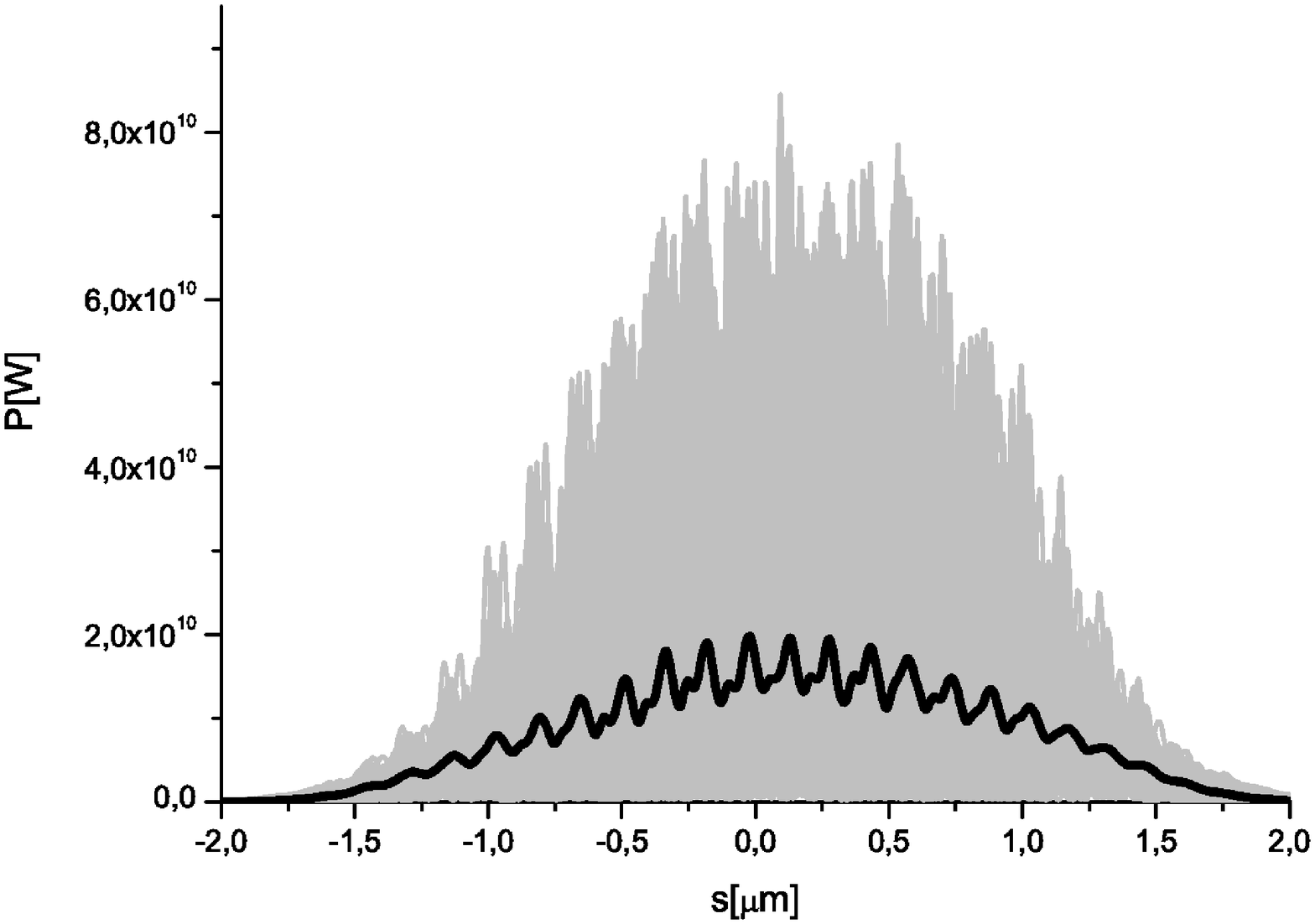}
\includegraphics[width=0.5\textwidth]{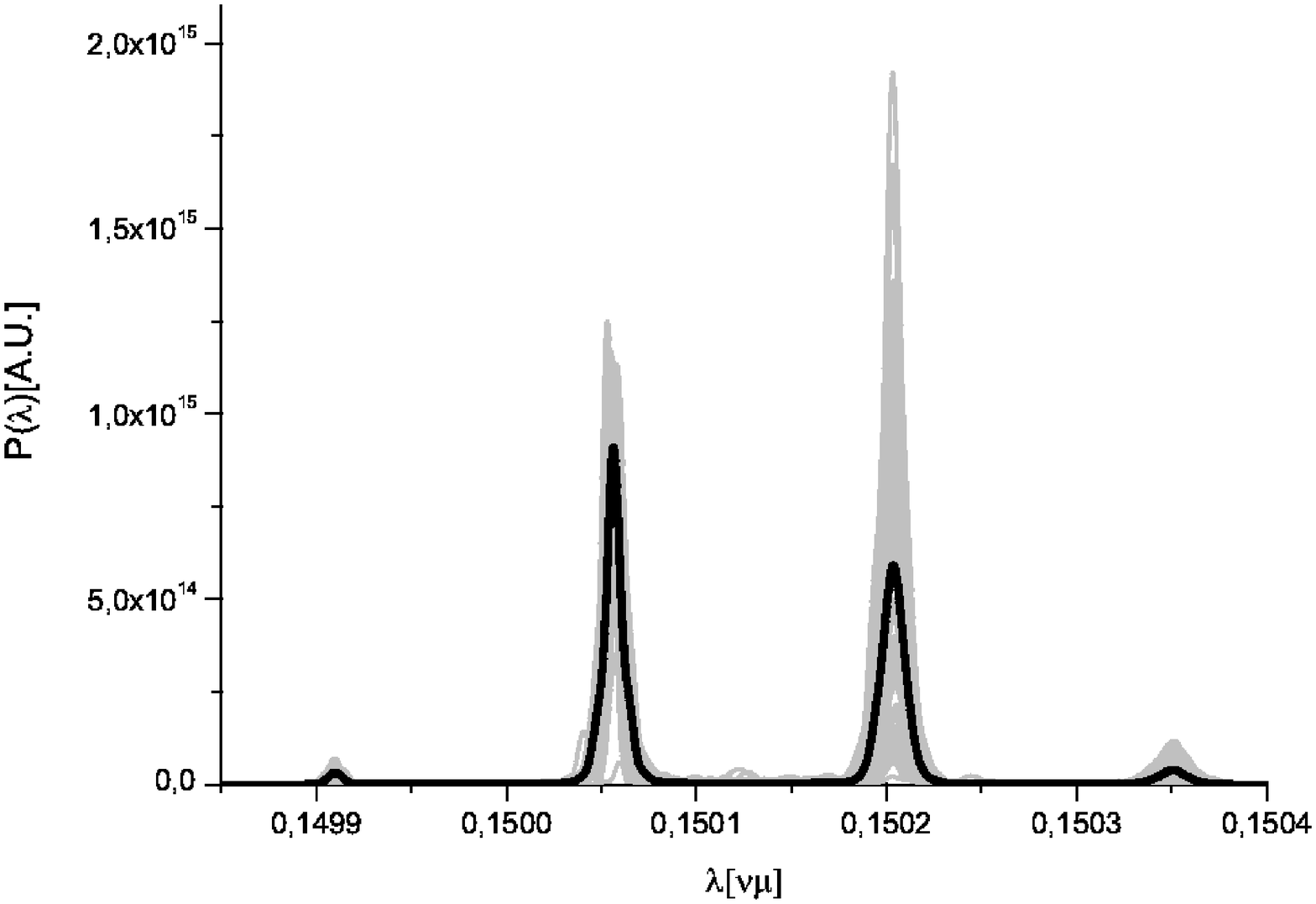}
\caption{(left) Power distribution of the X-ray radiation pulse at saturation.  (right) Spectrum  of the X-ray radiation pulse.  Grey lines refer to single shot realizations, the black line refers to the average over a hundred realizations.} \label{out}
\end{figure}

\begin{figure}[tb]
\includegraphics[width=0.5\textwidth]{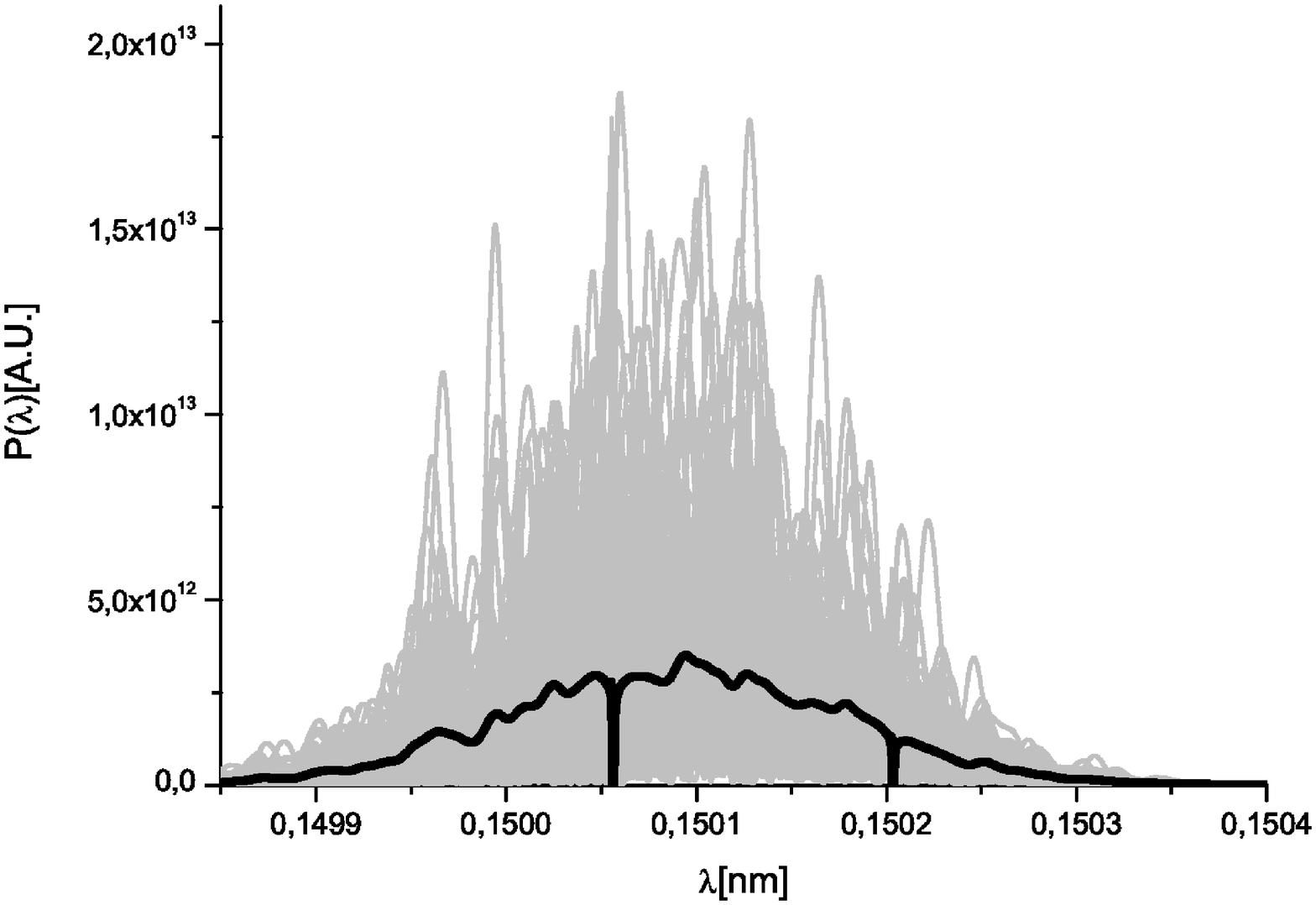}
\includegraphics[width=0.5\textwidth]{spout1.eps}
\includegraphics[width=0.5\textwidth]{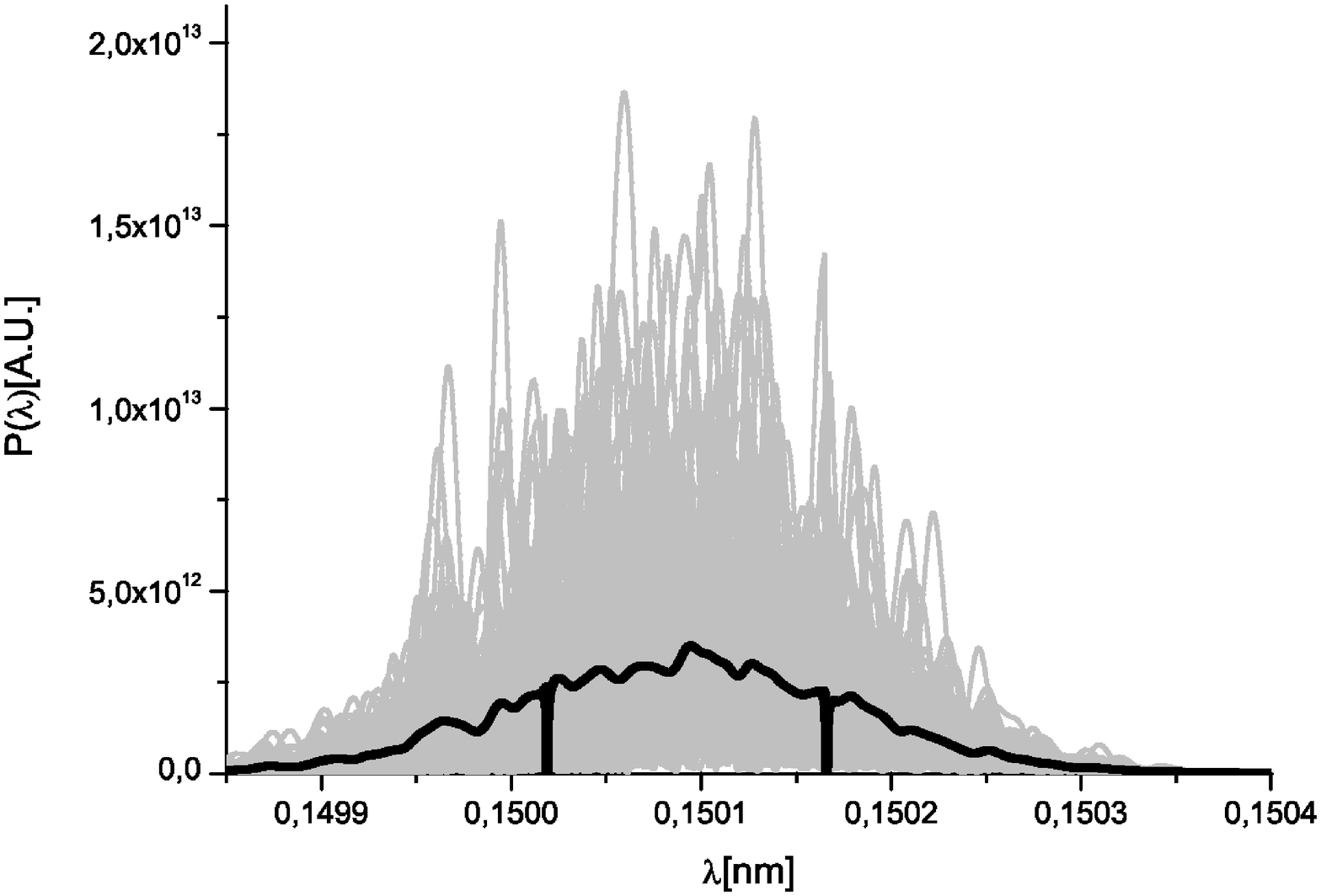}
\includegraphics[width=0.5\textwidth]{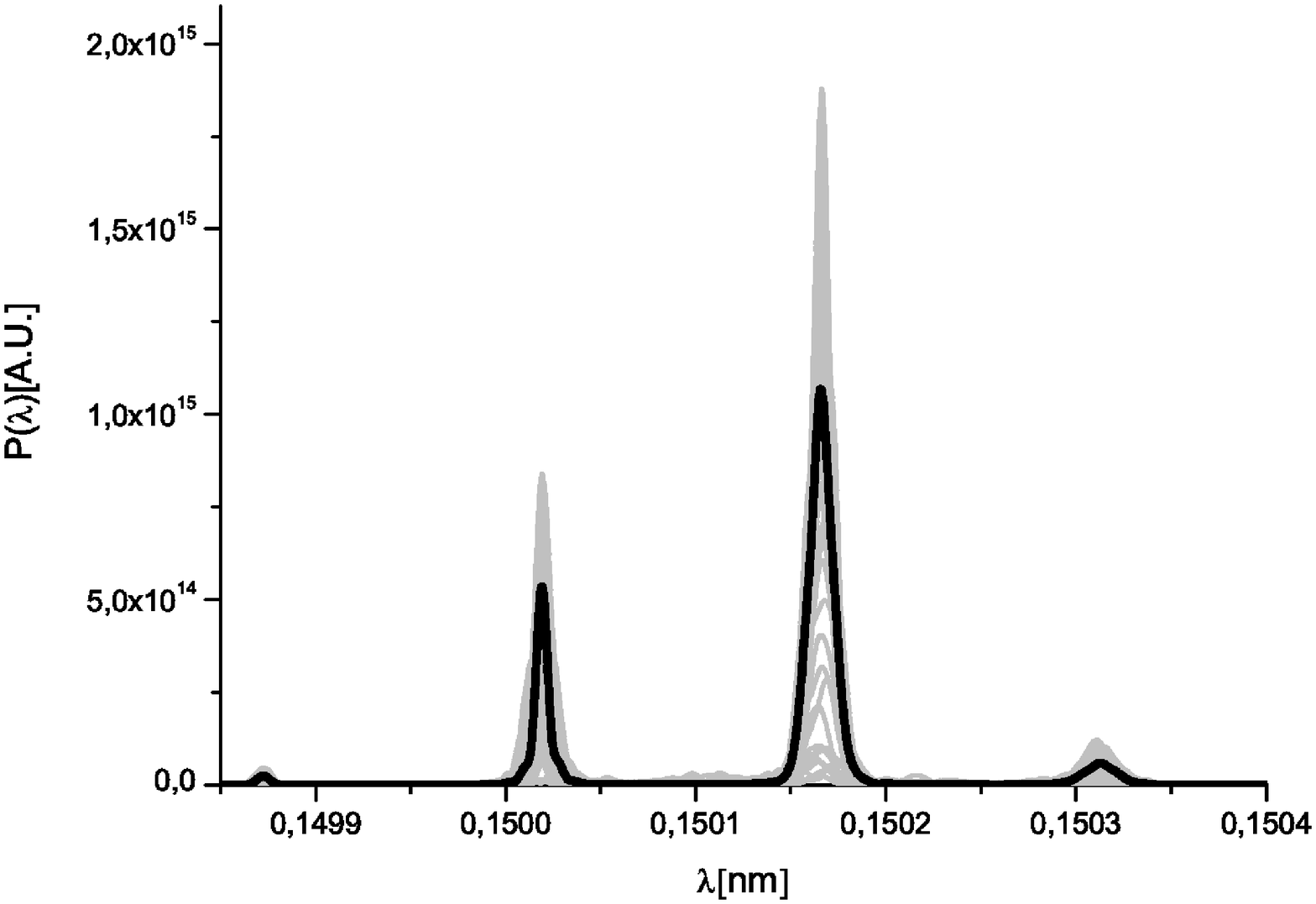}
\caption{Effect of different position of modes within the FEL gain band at fixed mode spacing. (left) Position of the frequency filters relative to the X-ray spectrum. (right) Spectrum  of the X-ray radiation pulse. Grey lines refer to single shot realizations, the black line refers to the average over a hundred realizations.} \label{outcmp}
\end{figure}

Photons and electrons are recombined after the crystals, so that the electron bunch temporally selects a part of the long tail in Fig. \ref{pseed}, and is seeded by it. The seeded electron beam goes through the output undulator, Fig. \ref{lclsd3}, which is sufficiently long (14 cells) to reach saturation. The output power and spectrum of the setup is shown in Fig. \ref{out}.
The effect of the double crystal is clear. The output power consists of a 20 GW pulse of coherent radiation, divided in two separate longitudinal modes, yielding doublet spectral lines on the right part of Fig. \ref{out}. By inspection of Fig. \ref{out} one can see that the power is not evenly distributed between the two modes. In particular, the low frequency mode is slightly dominant. The way the two modes compare with respect to each other depends in a complicated way on the position, in frequency, of the band-stop filters with respect to the impinging SASE spectrum, i.e. on the angular tilt of the crystals. Fig. \ref{outcmp} shows several simulation cases demonstrating different possible situations.

It should be noted that the phase of the sinusoidal envelope of the seed signal (see Fig. \ref{pseed} right) is randomly distributed from $0$ to $2 \pi$ on a shot-to-shot basis. As a result, the output power in Fig. \ref{out}, averaged over $100$ shots, does not exhibit any "mode-beat" component. Small residual oscillations, see Fig .\ref{out} left, are only due to the fact that the averaging process is performed on a finite number of shots (one hundred) only.

Additionally, it is interesting to remark that the average output power at saturation ($15$ GW) is about the same  that we had in the previous case \cite{OURY4}, where a single spectral line was generated. In other words here we only have a redistribution of the total number of photons between two modes, but the total number of photons remains about constant.

Finally, we note how the spectra in Fig. \ref{outcmp} right, exhibit small sidebands on the left, and on the right of the doublet. This effect is ascribed to nonlinear interaction of the two seed waves. This leads to generation of a discrete spectrum of beat waves extending beyond the FEL gain band. In particular, in the spectral window shown in the plots, we can distinguished two nonlinear resonances at $2 \omega_1- \omega_2$ and $2 \omega_2 - \omega_1$.  Generation of such wave beat resonance has been demonstrated  and studied in the case of LCLS parameters in \cite{FREU}.

\newpage

\section{\label{sec:COTR} Coherent OTR from an optically modulated electron bunch}

At the exit of the output undulator, the electron bunch exhibits substantial energy modulation at wavelengths in the range of the spacing between the doublet lines. This behavior, confirmed by simulations, see Fig. \ref{sampledat} bottom right is to be expected on the basis of qualitative reasoning as well. In fact, in the output undulator microbunching develops at both seeding wavelengths, leading to a beating with frequency related to the difference frequency. This beating is evident from the analysis of the output power from a given single-shot simulation, e.g. the one in Fig. \ref{sampledat}, upper right. It follows that electrons lose energy following such beating pattern, thus leading to the development of a strong energy modulation.

According to the scheme described in Section \ref{sec:oper}, Fig. \ref{lclsd5}, a weak dispersive element transforms such energy modulation in a substantial density modulation. A longitudinal dispersion of several tens microns is enough to optimize the modulation level. Starting with the output from Genesis, we performed a simulation of the bunching in the chicane with the help of the 3D tracking code Elegant \cite{ELEG}. The result is shown in Fig. \ref{sampledat}, bottom right plot, where we consider a case study with the beam current profile $I(t)$ after a chicane with $R_{56} = 0.08~$mm. Such chicane can be as short as a few meters. In this Section we estimate the number of optical photons emitted by the modulated electron bunch as it travels between the last bend of the chicane and the OTR screen, Fig. \ref{lclsd5}. We base our investigation on the analysis presented in \cite{OOTR}.

Let us define the slowly varying envelope of the field in the
space-frequency domain as $\vec{\widetilde{E}} = \vec{\bar{E}}
\exp{[-i \omega z/c]}$. We will refer to this quantity simply as
"the field". Here $\vec{\bar{E}}(\omega, \vec{r},z)$ is the
Fourier transform of the electric field $\vec{{E}}(t, \vec{r},z)$
in the space-time domain, according to the convention:
$\vec{\bar{E}}(\omega, \vec{r},z) = \int_{-\infty}^{\infty}
\vec{{E}}(t, \vec{r},z) \exp[i \omega t] d t$. Note that $\vec{r}$
indicates the transverse position vector.

The energy radiated by the bunch per unit frequency interval per unit surface can be calculated as in \cite{JACK}, and expressed in terms of number of photons, yielding:


\begin{eqnarray}
\frac{d N_\mathrm{ph}}{d\omega d S} =  \frac{c}{4\pi^2 \hbar \omega} \left|\vec{\widetilde{E}}(z,\vec{r})\right|^2~. \label{radden2}
\end{eqnarray}
In order to obtain the field produced by the electron bunch at the position of the OTR screen, $\vec{\widetilde{E}}(\vec{r})$, a microscopic approach can be used where the single-electron field is averaged over the six-dimensional phase-space distribution of electrons. In this (Lagrangian) approach particles are labeled with a given index, and the motion of individual charges is tracked through space. One follows the evolution of each particle as a function of energy deviation $\delta \gamma$, angular direction $\vec{\eta}$, position $\vec{l}$ and arrival time $t$ at a given longitudinal reference-position. Knowing the evolution of each particle, individual contributions to the field are separately calculated and summed up. Due to the high-quality electron beams produced at XFELs (highly collimated and nearly monochromatic) we have the simplest possible situation. Namely, when performing OTR calculations from an optically modulated electron bunch we can neglect both angular and energy distribution, and use a model of a cold electron bunch with given longitudinal, $f_\tau(t)$, and transverse, $f_l(\vec{l})$, charge density distributions. Note that in general we have no factorization of the charge density distribution into longitudinal $f_\tau(t)$ and transverse $f_l(\vec{l})$ factors. However, here we will be interested in an estimate of the number of available photons and, with some accuracy, we  can use a model with separable charge density distribution function. In fact, this assumption is not related to fundamental principles, and will only lead to a different numerical factor in the estimation of the number of photons.

Let us first consider the transverse charge density distribution $f_l(\vec{l})$. In the following we will simply assume a Gaussian shape:

\begin{eqnarray}
f_l(\vec{l}) = \frac{1}{\sqrt{2\pi} \sigma_r} \exp\left[-\frac{l^2}{2 \sigma_r^2}\right]
\label{fl}
\end{eqnarray}
Let us now turn to the longitudinal charge density distribution $f_\tau(t)$. In our study case, the electron beam exhibits a relatively broadband modulation around the nominal wavelength of $200$ nm. It is therefore convenient to write $f_\tau(t)$ as

\begin{eqnarray}
f_\tau(t) = f_{\tau 0}(t) + f_{\tau 1}(t)
~.\label{ftautau}
\end{eqnarray}
Here $f_{\tau 0}$ is related to the unmodulated electron beam current by

\begin{eqnarray}
f_{\tau 0}(t) = \frac{1}{\sqrt{2\pi} \sigma_T} \frac{I_0(t)}{I_\mathrm{max}} = \frac{1}{\sqrt{2\pi} \sigma_T} \exp\left[-\frac{t^2}{2\sigma_T^2}\right]~,
\label{ftau0}
\end{eqnarray}

\begin{figure}[tb]
\includegraphics[width=1.0\textwidth]{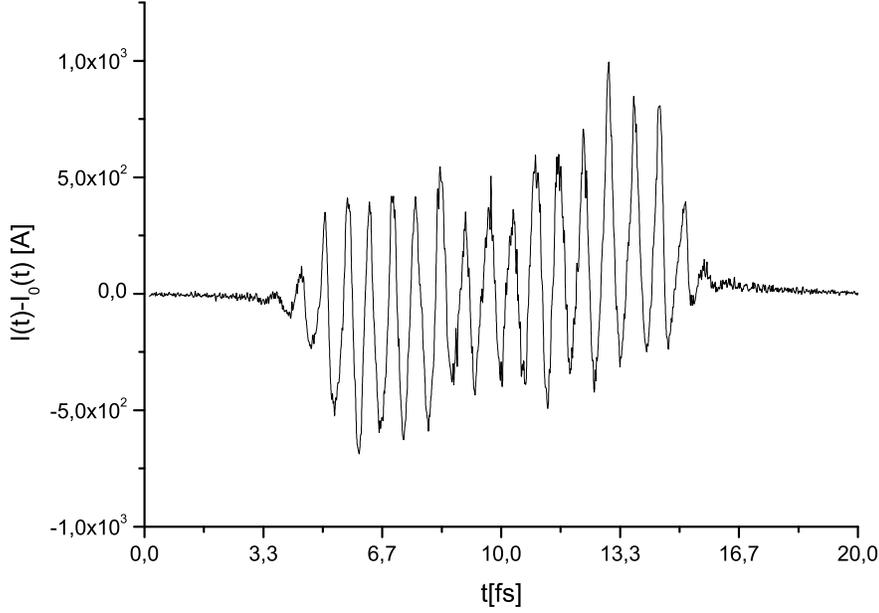}
\caption{$I(t)-I_0(t)$ as a function of time. This plot shows that part of the electron beam current responsible for electron beam density modulation at optical wavelengths.} \label{ftau1}
\end{figure}

where $I_\mathrm{max} = 3$ kA, and $\sigma_T \simeq 3$ fs is the electron bunch rms duration, which we assumed to be Gaussian throughout all this paper. Using a similar definition, the modulation part of the charge density distribution, $f_{\tau 1}(t)$ can be found as

\begin{eqnarray}
f_{\tau 1}(t) = \frac{I(t)-I_0(t)}{\sqrt{2\pi} \sigma_T I_\mathrm{max} }~,
\label{ftau1}
\end{eqnarray}
and is explicitly shown in Fig. \ref{ftau1} for the numerical case under investigation.

Having said this, the field distribution for the electron bunch at the OTR screen in the space-frequency domain is essentially a convolution in the space domain of the temporal Fourier transform of the charge density distribution and the temporal Fourier transform of the single-electron field, yielding

\begin{eqnarray}
\vec{\widetilde{E}}(\vec{r}) = N_e \bar{f}_\tau(\omega) \int d
\vec{r'} \vec{\widetilde{E}}^{(1)}(\vec{r'})
f_l(\vec{r}-\vec{r'})~,\label{totalf0}
\end{eqnarray}
where $\bar{f}_\tau(\omega)$ is the Fourier transform of the
temporal charge density distribution ${f}_\tau(t)$, and $\vec{\widetilde{E}}^{(1)}$ is the electric field from a single electron.

Since we need to calculate the total number of OTR photons available, the longitudinal position of the observation plane, $z$, does not matter, and it is easier to perform calculations in the far zone for calculation purposes only. From Eq. (\ref{totalf0}), from the definition of spatial Fourier transform, and Eq. (\ref{radden2}) we obtain

\begin{eqnarray}
\frac{d N_\mathrm{ph}}{d\omega d S} = N_e^2 \frac{d N^{(1)}_\mathrm{ph}}{d\omega d S} \left|\varrho(\omega, \vec{\xi})\right|^2~, \label{radden2}
\end{eqnarray}

where $N_e^2$ is the number of electrons in the bunch, $(1)$ refers to a single electron,  and

\begin{eqnarray}
\varrho\left(\omega ,\vec{\xi}\right) =
\bar{f}_\tau(\omega)\bar{f}_l(\vec{\xi})~,\label{bigf2}
\end{eqnarray}
$\bar{f}_l(\vec{\xi})$ being the Fourier transform of transverse charge density distribution, with $\vec{\xi}$ the angular variable conjugated to $\vec{r}$. Let us study each of the factors in Eq. (\ref{radden2}) separately.

\begin{figure}[tb]
\includegraphics[width=1.0\textwidth]{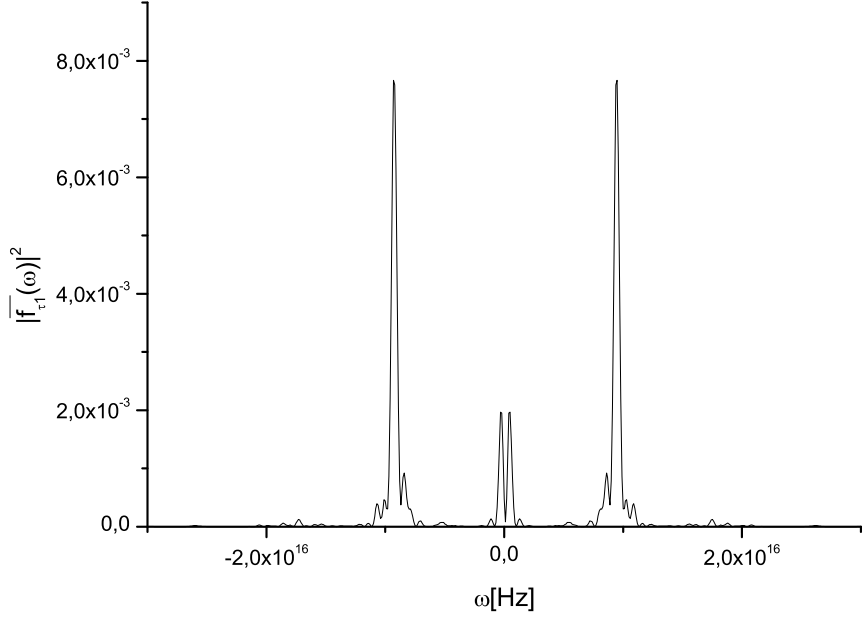}
\caption{Squared modulus of the Fourier transform of $f_{\tau1}$, $|\bar{f}_{\tau1}|^2$, as a function of the frequency.} \label{Lbarftau1}
\end{figure}
The spatial Fourier transform of $f_l$, i.e. $\bar{f}_l$ may simply be written as

\begin{eqnarray}
\bar{f}_{l}(\xi) = \exp\left[-\frac{\omega \xi^2 \sigma_r^2}{2 c^2}\right]~,
\label{ftfl}
\end{eqnarray}
which is dimensionless.  $\bar{f}_\tau$ can  be calculated as the sum of two parts:

\begin{eqnarray}
\bar{f}_\tau(\omega) = \bar{f}_{\tau 0}(\omega) + \bar{f}_{\tau 1}(\omega)~,
\label{ftftau}
\end{eqnarray}
where

\begin{eqnarray}
\bar{f}_{\tau 0}(\omega) = \exp\left[-\frac{\sigma_T^2 \omega^2 }{2}\right]
\label{ftftau0}
\end{eqnarray}
and $\bar{f}_{\tau 1}$ is calculated numerically,  $|\bar{f}_{\tau 1}|^2$ being shown in Fig. \ref{Lbarftau1}.

Finally, the photon density distribution from a single electron can be calculated from the knowledge of the field angular distribution relative to a single electron according to:

\begin{eqnarray}
\frac{d N_\mathrm{ph}^{(1)}}{d\omega d S} =  \frac{c}{4\pi^2 \hbar \omega} \left|\vec{\widetilde{E}}^{(1)}(z,\vec{\xi})\right|^2~. \label{radden2b}
\end{eqnarray}
Following \cite{OOTR}, one obtains the far-zone expression for the single-particle field

\begin{eqnarray}
\vec{\widetilde{E}}\left({z},\vec{\xi}\right) =  \frac{2 e
\gamma^2\vec{{\xi}}}{c {z} (\gamma^2 \xi^2 + 1)} \exp\left[\frac{i
\omega  \xi^2}{2 c }z\right] ~. \label{Efarsum3}
\end{eqnarray}
As said before, Eq. (\ref{Efarsum3}) is the spatial Fourier transform of the field at the OTR screen:

\begin{eqnarray}
\vec{\widetilde{E}}\left(\vec{{r}}\right) &=&  - \frac{2  \omega
e}{c^2 \gamma} \frac{\vec{r}}{{r}} K_1\left(\frac{\omega{r}}{c
\gamma} \right)~,\label{vir4ea}
\end{eqnarray}
where $K_1$ is the first order modified Bessel function of the second kind. Eq. (\ref{vir4ea}) is valid, strictly speaking, only in the region $r \ll \gamma \lambdabar$ on the OTR screen. However, as noted in \cite{OOTR}, corrections to Eq. (\ref{vir4ea}) enter only logarithmically in the calculation of the number of photons outside such region. Therefore, as concerns photon number estimations, we may still use Eq. (\ref{vir4ea}), yielding Eq. (\ref{Efarsum3}) which in its turn, substituted in Eq. (\ref{radden2b}), gives

\begin{eqnarray}
\frac{d N_\mathrm{ph}^{(1)}}{d\omega d S} = \frac{ \alpha}{\pi^2
\omega {z}^2} \frac{\gamma^4
\xi^2}{(\gamma^2\xi^2+1)^2}~,\label{radden2c}
\end{eqnarray}
with $\alpha \equiv e^2/(\hbar c) = 1/137$ the fine structure
constant.

Substitution of Eq. (\ref{radden2c}) and Eq. (\ref{bigf2}) in Eq. (\ref{radden2}) should finally be followed by integration over a proper bandwidth and angular range to yield the number of photon available.

\begin{figure}[tb]
\includegraphics[width=1.0\textwidth]{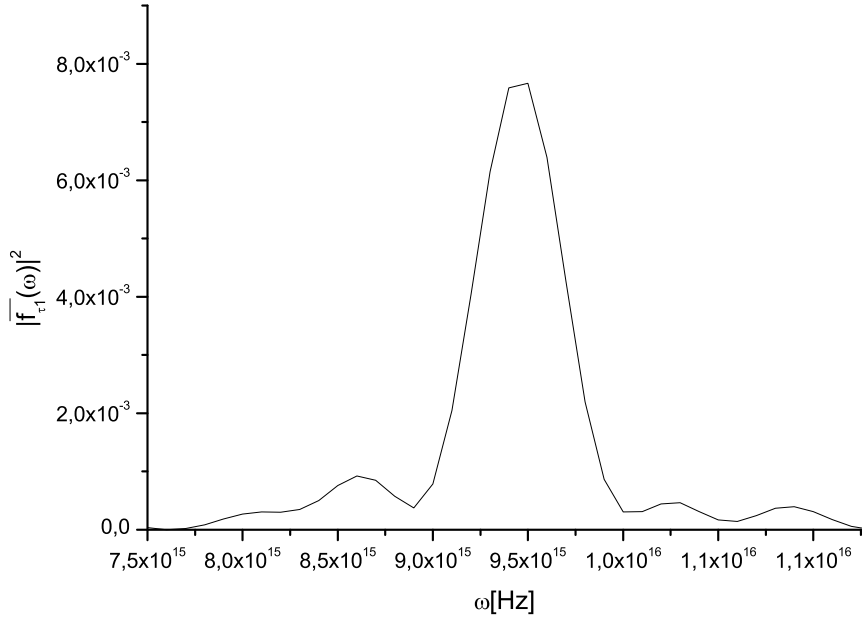}
\caption{Squared modulus of the Fourier transform of $f_{\tau1}$, $|\bar{f}_{\tau1}|^2$ in the range between $167$nm and $251$nm.} \label{barftau1}
\end{figure}
Application, with some accuracy, of the adiabatic approximation, meaning that the bunch duration $\sigma_T$ is much larger than the period of the modulation, allows one to restrict attention to a small bandwidth around the main modulation wavelength at $200$ nm. Effectively, we choose a bandwidth between $167$ nm and $251$ nm, corresponding to the spectrum $\bar{f}_{\tau 1}$ in Fig. \ref{barftau1}, which is a selection of Fig. \ref{barftau1}. Finally, we assume an optics with large numerical aperture $0.1$ yielding about $10^{11}$ photons. Such number is more than enough for the purposes considered in Section \ref{sec:oper}.

\section{\label{conc} Conclusions}

In this article we presented an extension to our previously proposed method for self-seeding \cite{OURX}-\cite{OURY4}, which was based on the use of a single Diamond crystal in Bragg geometry acting as a band-stop filter. Here we extend such treatment to a series of two or more crystals tuned at closely-spaced frequencies, and we demonstrate the possibility of generating doublet spectral lines through the self-seeding process. In other words, fully coherent radiation is shared between two longitudinal modes. Depending on the crystals settings one can have one mode dominating on the other, or a balance between the two\footnote{It should be noted that the idea of radiation shared between longitudinal modes should be thought of from a statistical standpoint. In other words, here we are talking about an average over ensemble of shots, not about a single realization.}. Moreover, the spacing between the two modes can be tuned within the FEL gain band, i.e. within $10$ eV. Once doublet spectral lines are generated, they can be used to study processes dealing with a large change in cross-section over a narrow wavelength range.

Furthermore, possibility of pump-probe experiments with sub-femtosecond resolution where an external optical laser is involved are enabled. In fact, a power beating between the two closely spaced wavelengths of the doublet yields a modulation of the SASE output power on the visible range. By energy conservation, such modulation is printed on the electron bunch as an energy modulation at the same frequencies, which can be transformed into a density modulation with the help of a weak dispersive element. The modulated electron bunch can then be made radiating coherently with the help of any radiator, e.g. an OTR screen. A powerful pulse of coherent radiation in the optical range is then produced, which is intrinsically synchronized to the X-ray photon pulse, and can be cross-correlated to the external pump-laser. Such correlation results in the possibility of measuring the jitter between SASE doublet and external optical pump with a few femtosecond accuracy. The coherent optical pulse may also be used to monitor the formation of the doublet, since it will appear only when the doublet is actually present, which enables beating.

From the technical viewpoint, the present method constitutes a minor upgrade of our previously proposed technique, of which it retains flexibility, simplicity of implementation, low cost and no risk for the recovering of the baseline mode of operation.

\section{Acknowledgements}

We are grateful to Massimo Altarelli, Reinhard Brinkmann, Serguei
Molodtsov and Edgar Weckert for their support and their interest
during the compilation of this work.

\end{document}